# A Model of Two-Way Selection System for Human Behavior

**Bin Zhou[1,2], Shujia Qin[3], Xiao-Pu Han[4], Zhe He[1], Jia-Rong Xie[1], Bing-Hong Wang[1,5]***

**1** Department of Modern Physics and Nonlinear Science Center, University of Science and Technology of China, Hefei, China, **2** Suzhou Institute of Technology, Jiangsu University of Science and Technology, Suzhou, China, **3** State Key Lab of Robotics, Shenyang Institute of Automation, Chinese Academy of Sciences, Shenyang, China, **4** Institute of Information Economy and Alibaba Business College, Hangzhou Normal University, Hangzhou, China, **5** College of Physics and Electronic Information Engineering, Wenzhou University, Wenzhou, China

## Abstract

Two-way selection is a common phenomenon in nature and society. It appears in the processes like choosing a mate between men and women, making contracts between job hunters and recruiters, and trading between buyers and sellers. In this paper, we propose a model of two-way selection system, and present its analytical solution for the expectation of successful matching total and the regular pattern that the matching rate trends toward an inverse proportion to either the ratio between the two sides or the ratio of the state total to the smaller group's people number. The proposed model is verified by empirical data of the matchmaking fairs. Results indicate that the model well predicts this typical real-world two-way selection behavior to the bounded error extent, thus it is helpful for understanding the dynamics mechanism of the real-world two-way selection system.

**Funding:** This work was funded by the National Important Research Project (Grant No. 91024026), the National Natural Science Foundation of China (No. 11205040, 11105024, 11275186), the Major Important Project Fund for Anhui University Nature Science Research (Grant No. KJ2011ZD07) and the Specialized Research Fund for the Doctoral Program of Higher Education of China (Grant No. 20093402110032). The funders had no role in study design, data collection and analysis, decision to publish, or preparation of the manuscript.

**Competing Interests:** The authors have declared that no competing interests exist.

* E-mail: bhwang@ustc.edu.cn

## Introduction

Human-initiated systems always run in a complex way. In the past ten years, related work mainly focused on the temporal and spatial distribution characteristics of human activity patterns. Because of the complexity of human behavior, many underlying mechanisms have not been discovered yet. The two-way selection scenario among humans is one of the complicated but common phenomena in daily life. It happens in the processes like choosing a mate between men and women, making contracts between job hunters and recruiters, and trading between buyers and sellers. In a sense, two-way selections can be regarded as the base of building many social relationships. Generally, the participants in a two-way selection process are first classified into two groups by their natural status. Then they observe, study the factors of the people on the other side, and finally make their choices. For instance, in the case of marriages, one's appearance, personality, wealth, and sense of humor, are prevalently taken into consideration. Besides the individual characters, impersonal factors also exert an influence, e.g. the member totals on each side and their ratio. How many characters will be inspected and chosen deeply affects the result of a selection process. However, usually it is difficult to compare and to distinguish these characteristics quantitatively even qualitatively through traditional methods, such as psychological tests and social surveys.

The well-known marriage game in statistical physics has been researched in these papers [1–6], whose main novel concept is the stability of marriages. This view point aims to find a stable matching between the two sets of men and women. Such a model results in the destiny that every one in the sets gets married and the final marriage relationships are "stable". However, the internal mechanism of a two-way selection system can be modeled in another way: not all of the participants have to get married in one trial of the processes, i.e. some of them would be successful in matching but the others not. This mechanism would render assistance to some social problems, such as the prediction of the total of friendships or other gregarious relations [7–15]. In this paper, we present a model for two-way selections to investigate the factors influencing the matching rate. The data of matchmaking fairs are analyzed to support our model. Based on this model, the method of estimating the number of factors impacting people's decisions is also proposed.

## The Model and Analytical Results

Our model of the two-way selection is stated as follows:

i) The system has two sets of agents, $A$ and $B$, respectively amounting to $k_1$ and $k_2$.

ii) The $i$th agent in set $A$ (or set $B$) has its own character denoted by $c_{A_i}$ (or $c_{B_i}$). Correspondingly, the character the $i$th agent attempts to select is denoted by $s_{A_i}$ (or $s_{B_i}$).

iii) The agents' characters are denoted by integers without loss of generality. Assume the characters has $n$ types, i.e. $c_{A_i}$, $c_{B_j}$, $s_{A_i}$, $s_{B_j} \in \Omega$, $\Omega = \{1,2,\ldots,n\}$, $i \in \{1,2,\ldots,k_1\}$, $j \in \{1,2,\ldots,k_2\}$. In one trial of the model, $c_{A_i}$, $c_{B_j}$, $s_{A_i}$, $s_{B_j}$ pick an element in $S$ following the uniform distribution.

iv) The condition of successful matching of two agents $A_i$ and $B_j$ is $c_{A_i} = s_{B_j}$ and $s_{A_i} = c_{B_j}$. That is, when agent $A_i$'s character meets agent $B_j$'s requirement and *vice versa*, agent $A_i$ and agent $B_j$ have a successful matching.

For given $k_1$, $k_2$ and $n$, the expectation $E$ of the total number of matching pairs in the model is

$$E(k_1,k_2,n)= k_1 \sum_{i=0}^{k_1}\left[\sum_{j=i+1}^{k_2} C_{k_1-1}^{i} C_{k_2}^{j}\left(\frac{1}{n^2}\right)^{i+j}\left(\frac{n^2-1}{n^2}\right)^{k_1+k_2-i-j-1}\right] + k_2 \sum_{j=0}^{k_2}\left[\sum_{i=j}^{k_1} C_{k_1}^{i+1} C_{k_2-1}^{j-1}\left(\frac{1}{n^2}\right)^{i+j}\left(\frac{n^2-1}{n^2}\right)^{k_1+k_2-i-j-1}\right]. \quad (1)$$

According to

$$\lim_{k\to\infty} C_k^i x^i (1-x)^{k-i} = \frac{e^{-\lambda}\lambda^i}{i!},$$
$$\sum_{i=0}^{\infty} \frac{x^i}{(i!)^2} = \mathrm{BesselI}(0,2\sqrt{x}),$$
$$\sum_{i=0}^{\infty} \frac{x^i}{(i!)(i-1)!} = \sqrt{x}\,\mathrm{BesselI}(1,2\sqrt{x}),$$

where $\lambda = xk$, $\mathrm{BesselI}(0,2\sqrt{x})$ and $\mathrm{BesselI}(1,2\sqrt{x})$ are the modified Bessel functions of the first kind, the expectation in (1) can approximate to

$$E \approx k_1\left[1 - e^{-\lambda_1-\lambda_2}\left(\mathrm{BesselI}\left(0,2\sqrt{\lambda_1\lambda_2}\right) + \sqrt{\frac{\lambda_1}{\lambda_2}}\mathrm{BesselI}\left(1,2\sqrt{\lambda_1\lambda_2}\right)\right)\right], \quad (2)$$

where $\lambda_1 = k_1/n^2$, $\lambda_2 = k_2/n^2$.

Due to the symmetry of $k_1$ and $k_2$ in (1), without loss of generality, we just study the $k_1 \le k_2$ case under three conditions: $k_1 \le k_2 \ll n^2$, $k_1 \le n^2 \ll k_2$, and $n^2 \ll k_1 \le k_2$. When $k_1 \le k_2 \ll n^2$, resulting in $\lambda_1 \to 0, \lambda_2 \to 0$, calculating the zeroth power term and the first power term of (2) obtains

$$E \approx \frac{1}{n^2} k_1 k_2. \quad (3)$$

When $k_1 \le n^2 \ll k_2$ or $n^2 \ll k_1 \le k_2$, according to

$$\lim_{x\to\infty} \mathrm{BesselI}(0,2x) = \lim_{x\to\infty} \mathrm{BesselI}(1,2x) = \frac{e^{2x}}{2\sqrt{\pi x}}, \quad (4)$$

Equation (2) can be simplified as

$$E \approx k_1 - k_1 \frac{e^{-(\sqrt{\lambda_1}-\sqrt{\lambda_2})^2}}{2\sqrt{\pi\sqrt{\lambda_1\lambda_2}}}\left(1+\sqrt{\frac{\lambda_1}{\lambda_2}}\right). \quad (5)$$

Because in this case $k_2$ is very large, Equation (5) can be further simplified as

$$E \approx k_1. \quad (6)$$

Define

$$\eta = \frac{k_2}{k_1}, \quad \xi = \frac{n^2}{k_1}, \quad P = \frac{2E}{k_1+k_2}, \quad (7)$$

where $\eta$ denotes the ratio of $k_2$ to $k_1$; $\xi$ denotes the ratio of $n^2$ to $k_1$; $P$ denotes the estimated ratio of successful matching pairs to the average number of two type agents. Then Equation (2) can be transformed into

$$P \approx \frac{2}{1+\eta}\left[1 - e^{-\frac{1}{\xi}-\frac{\eta}{\xi}}\left(\mathrm{BesselI}\left(0,\frac{2\sqrt{\eta}}{\xi}\right) + \frac{1}{\sqrt{\eta}}\mathrm{BesselI}\left(1,\frac{2\sqrt{\eta}}{\xi}\right)\right)\right]. \quad (8)$$

Equation (3) can be written as:

$$P \approx \frac{2\eta}{\xi(\eta+1)}, \quad 1 \le \eta \ll \xi. \quad (9)$$

Equation (6) can be written as:

$$P \approx \frac{2}{\eta+1}, \quad 1 \le \xi \ll \eta \;\; or \;\; \xi \ll 1 \le \eta. \quad (10)$$

Figure 1 shows the comparison between the analytical predictions of (1) and the simulation results. Figure 2(a) shows the comparison between the analytical predictions of (9) and the simulation results under the condition $1 \le \eta \ll \xi$ and displays a power-law relation with the exponent $-1$ between $P$ and $\xi$. Figure 2(b) shows the comparison between the analytical predictions of (10) and the simulation results under the condition $1 \le \xi \ll \eta$ and displays a power-law relation with the exponent $-1$ between $P$ and $\eta$; Figure 2(d) shows the comparison between the analytical predictions of (10) and the simulation results under the condition $\xi \ll 1 \le \eta$ and displays the same power-law relation to the result in Figure 2(b). The above analytical predictions and simulation results are consistent with each other. That is to say all analytical results are reliable.

Consider a special case $k_1 = k_2 = k$, resulting in $\eta = 1$. On the one hand, Equation (9) can be simplified as

$$P \approx \frac{1}{\xi}. \quad (11)$$

The relation between $P$ and $\xi$ approximates a power law with the exponent $-1$, and this case is shown in Figure 2(c). Equation (10) can be simplified as $P \approx 1$, suggesting that almost all of the agents can match successfully under the condition $\xi \ll 1 \le \eta$. On the other hand, because the condition $k_1 = k_2 = k$ results in $\lambda_1 = \lambda_2$, from (5) we can obtain

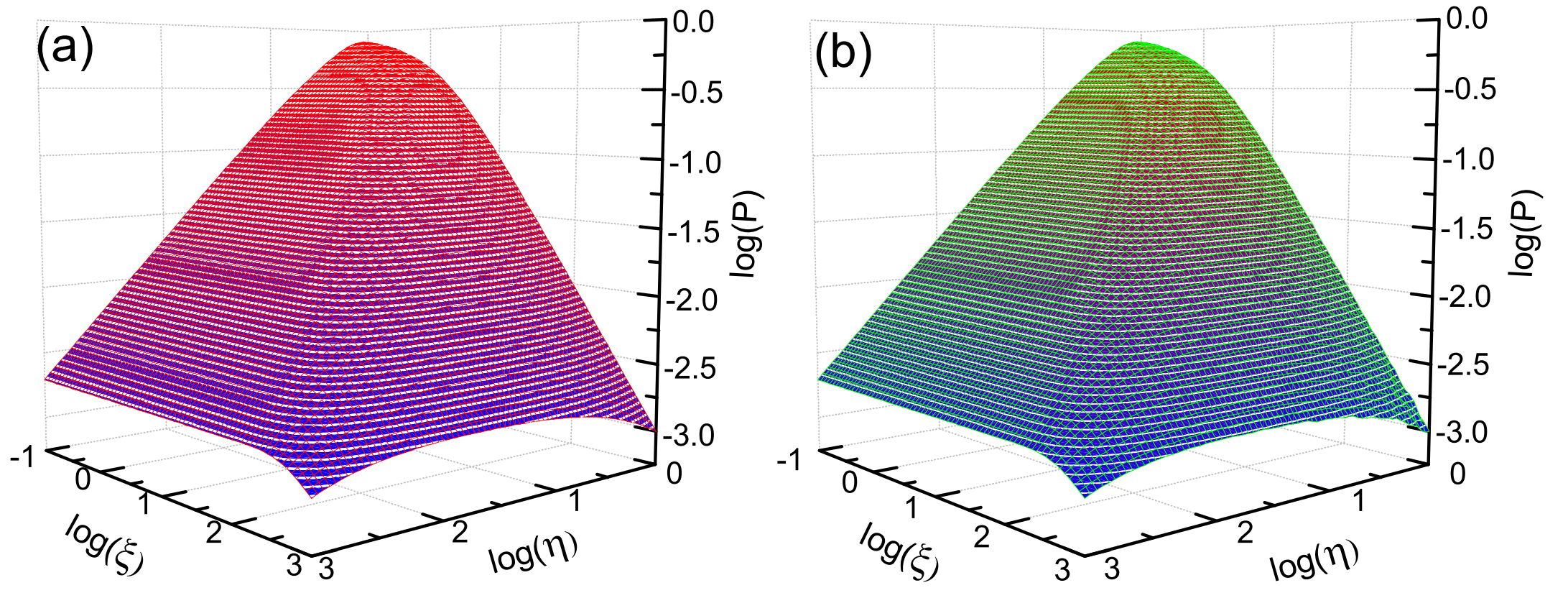

**Figure 1. The comparison between the analytical predictions and the simulation results.** In the two sub-figures, the parameter $k_1 = 100$; $\eta$ is assigned to values from 1 to 1000; $\xi$ is assigned to values from 0.1 to 1000. (a) shows analytical predictions of (1); (b) shows the simulation results.

$$E \approx k - \sqrt{\frac{n^2}{\pi}}\sqrt{k}. \qquad (12)$$

The second term of (12) is the number of the agents that can not successfully match in type A or type B. The larger $k$ is, the smaller $\sqrt{n^2/\pi}\sqrt{k}/k$ is. In reality, this is a result of fluctuation. The total combinations of the "own" state and the "expecting" state for an agent have $n^2$ possibilities in the model. In theory, the expected times of each state appearance is $k/n^2$. However, due to the fluctuations, almost all frequencies of every state appearance deviate around $k/n^2$. As a result, some agents can not successfully match. The number of times that each state may appear obeys the binomial distribution. The fluctuation is closely related to the standard deviation, according to the binomial theorem and standard deviation formula, we can obtain that the standard deviation equals $\sqrt{k(1-1/n^2)/n^2}$, which is directly proportional to $\sqrt{k}$. Thus, the number of the agents that can not successfully match is also proportional to $\sqrt{k}$. It explains the relationship between the second item of (12) and $\sqrt{k}$. From (12), we know that the proportionality coefficient is $\sqrt{n^2/\pi}$.

## The Verification Between the Model and Experimental Data

As the mate choosing between men and women is a typical real-world two-way selection system, eighty-two reported records of matchmaking fairs are analyzed to verify our model. Due to the uncertainty of approximation in these reports, we classify the data

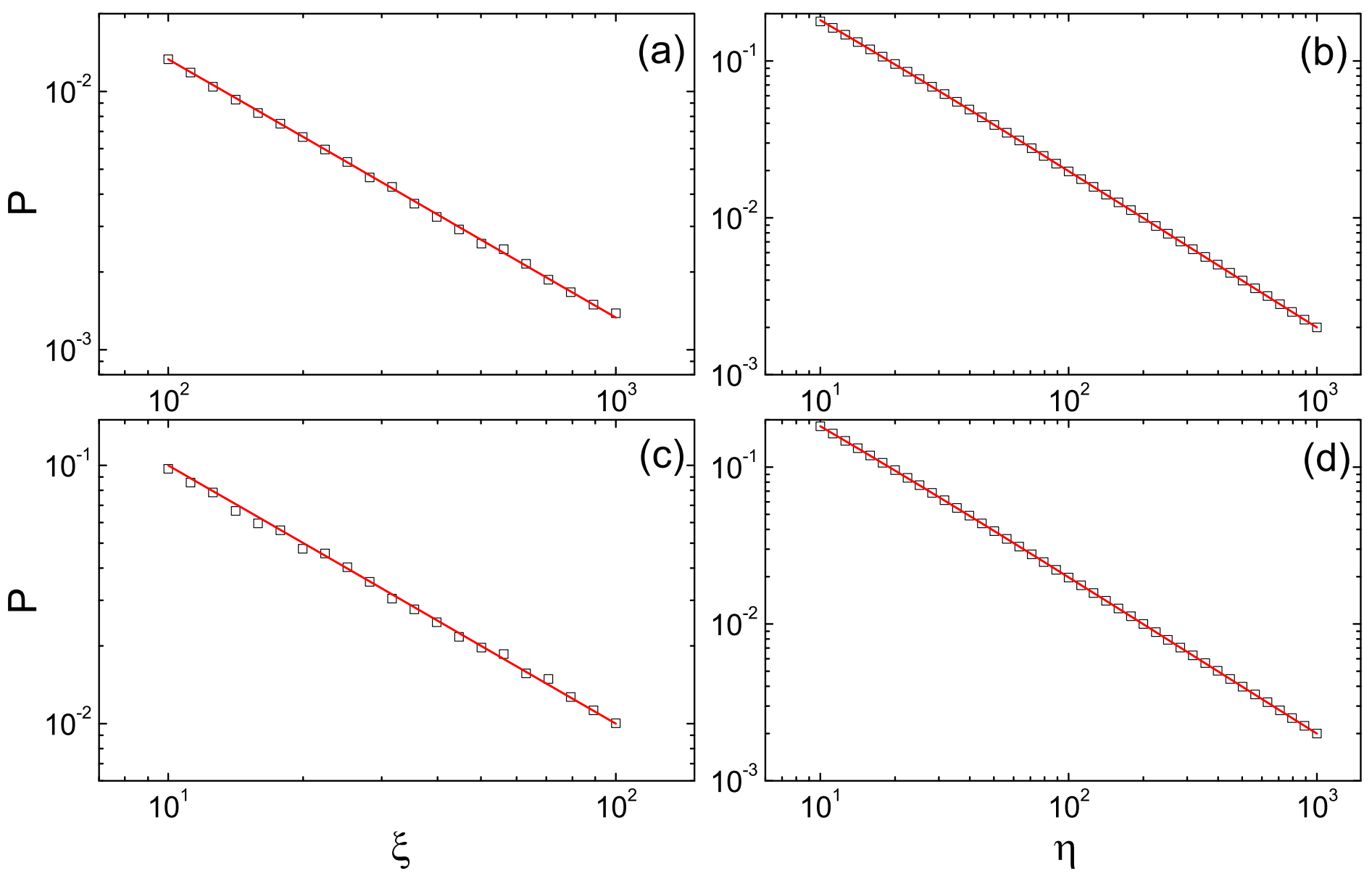

**Figure 2. The comparison between analytical predictions and the simulation results in the log-log plots.** In the four sub-figures, the parameter $k_1 = 100$; the squares are the simulation data; the solid lines are analytical predictions. In (a), $\eta = 2$; $\xi$ is assigned to values from 100 to 1000; the solid line is obtained from (9). In (b), $\xi = 2$; $\eta$ is assigned to values from 10 to 1000; the solid line is obtained from (10). In (c), $\eta = 1$; $\xi$ is assigned to values from 10 to 100; the solid line is obtained from (11). In (d), $\xi = 0.1$; $\eta$ is assigned to values from 10 to 1000; the solid line is obtained from (10).

**Table 1.** The data of matchmaking fairs.

| Website no. | Original descriptions | | Total participants $K$ | Matched pairs $E$ | Matching ratio $P$ |
|---|---|---|---|---|---|
| | joined | matched | | | |
| 01 | 13 | 3 | 13 | 3 | 0.4615 |
| 02 | 18 | 6 | 18 | 6 | 0.6667 |
| 03 | 20 | 1 | 20 | 1 | 0.1000 |
| 04 | 21 | 6 | 21 | 6 | 0.5714 |
| 05 | 25 | 5 | 25 | 5 | 0.4000 |
| 06 | 26 | 3 | 26 | 3 | 0.2308 |
| 07 | 26 | 4 | 26 | 4 | 0.3077 |
| 08 | 30 | 2 | 30 | 2 | 0.1333 |
| 09 | 32 | 4 | 32 | 4 | 0.2500 |
| 10 | 36 | 6 | 36 | 6 | 0.3333 |
| 11 | 36 | 6 | 36 | 6 | 0.3333 |
| 12 | 38 | 6 | 38 | 6 | 0.3158 |
| 13 | 40 | 8 | 40 | 8 | 0.4000 |
| 14 | >40 | 3 | 42±2 | 3 | 0.1432±0.0068 |
| 15 | >50 | 3 | 53±3 | 3 | 0.1136±0.0064 |
| 16 | ≈60 | 5 | 60±3 | 5 | 0.1671±0.0084 |
| 17 | >60 | 4 | 63±3 | 4 | 0.1273±0.0061 |
| 18 | >60 | 5 | 63±3 | 5 | 0.1591±0.0076 |
| 19 | >60 | ≈10 | 63±3 | 10±1 | 0.3197±0.0470 |
| 20 | 72 | 11 | 72 | 11 | 0.3056 |
| 21 | 80 | 5 | 80 | 5 | 0.1250 |
| 22 | 80 | 10 | 80 | 10 | 0.2500 |
| 23 | 80 | 13 | 80 | 13 | 0.3250 |
| 24 | 80 | 18 | 80 | 18 | 0.4500 |
| 25 | ~100 | 5 | 95±5 | 5 | 0.1056±0.0056 |
| 26 | 99 | 8 | 99 | 8 | 0.1616 |
| 27 | 100 | 5 | 100 | 5 | 0.1000 |
| 28 | ≈100 | 7 | 100±5 | 7 | 0.1404±0.0070 |
| 29 | >100 | 16 | 110±10 | 16 | 0.2933±0.0267 |
| 30 | 150 | 7 | 150 | 7 | 0.0933 |
| 31 | ~200 | 8 | 190±10 | 8 | 0.0844±0.0044 |
| 32 | ~200 | 22 | 190±10 | 22 | 0.2322±0.0122 |
| 33 | ≈200 | 4 | 200±10 | 4 | 0.0401±0.0020 |
| 34 | 206 | 10 | 206 | 10 | 0.0971 |
| 35 | >200 | 7 | 210±10 | 7 | 0.0668±0.0032 |
| 36 | >200 | 8 | 210±10 | 8 | 0.0764±0.0036 |
| 37 | >200 | 38 | 210±10 | 38 | 0.3627±0.0173 |
| 38 | 216 | 19 | 216 | 19 | 0.1759 |
| 39 | >240 | 22 | 252±12 | 22 | 0.1750±0.0083 |
| 40 | ≈258 | >10 | 258±13 | 11±1 | 0.0859±0.0121 |
| 41 | ~300 | 4 | 285±15 | 4 | 0.0282±0.0014 |
| 42 | ≈300 | 8 | 300±15 | 8 | 0.0535±0.0027 |
| 43 | >300 | >10 | 315±15 | 11±1 | 0.0703±0.0097 |
| 44 | >300 | 32 | 315±15 | 32 | 0.2036±0.0097 |
| 45 | 400 | ~20 | 400 | 19±1 | 0.0950±0.0050 |
| 46 | >500 | 3 | 525±25 | 3 | 0.0115±0.0005 |
| 47 | >500 | 8 | 525±25 | 8 | 0.0306±0.0015 |
| 48 | >500 | >10 | 525±25 | 11±1 | 0.0422±0.0058 |

**Table 1.** Cont.

| Website no. | Original descriptions | | Total participants $K$ | Matched pairs $E$ | Matching ratio $P$ |
|---|---|---|---|---|---|
| | joined | matched | | | |
| 49 | ~600 | ~40 | $570 \pm 30$ | 38±2 | $0.1341 \pm 0.0141$ |
| 50 | >600 | >78 | $630 \pm 30$ | 78 | $0.2482 \pm 0.0118$ |
| 51 | ~800 | 58 | $760 \pm 40$ | 58 | $0.1531 \pm 0.0081$ |
| 52 | >800 | >20 | $840 \pm 40$ | 21±1 | $0.0502 \pm 0.0047$ |
| 53 | ~1000 | ≈20 | $950 \pm 50$ | 20±1 | $0.0423 \pm 0.0043$ |
| 54 | ~1000 | 58 | $950 \pm 50$ | 58 | $0.1224 \pm 0.0064$ |
| 55 | ~1000 | 64 | $950 \pm 50$ | 64 | $0.1351 \pm 0.0071$ |
| 56 | ≈1000 | 12 | $1000 \pm 50$ | 12 | $0.0241 \pm 0.0012$ |
| 57 | ≈1000 | 15 | $1000 \pm 50$ | 15 | $0.0301 \pm 0.0015$ |
| 58 | ≈1000 | ~100 | $1000 \pm 50$ | 95±5 | $0.1910 \pm 0.0196$ |
| 59 | >1000 | 3 | $1050 \pm 50$ | 3 | $0.0057 \pm 0.0003$ |
| 60 | >1000 | 4 | $1050 \pm 50$ | 4 | $0.0076 \pm 0.0004$ |
| 61 | >1500 | 48 | $1575 \pm 75$ | 48 | $0.0611 \pm 0.0029$ |
| 62 | >1500 | >100 | $1575 \pm 75$ | 105±5 | $0.1339 \pm 0.0127$ |
| 63 | >1600 | 31 | $1680 \pm 80$ | 31 | $0.0370 \pm 0.0018$ |
| 64 | >2000 | ~100 | $2100 \pm 100$ | 95±5 | $0.0909 \pm 0.0091$ |
| 65 | >2000 | >113 | $2100 \pm 100$ | 119±6 | $0.1139 \pm 0.0111$ |
| 66 | ~3000 | ~100 | $2850 \pm 150$ | 95±5 | $0.0670 \pm 0.0070$ |
| 67 | ≈3000 | 186 | $3000 \pm 150$ | 186 | $0.1243 \pm 0.0062$ |
| 68 | >3000 | >200 | $3150 \pm 150$ | 210±10 | $0.1339 \pm 0.0127$ |
| 69 | >4000 | >500 | $4200 \pm 200$ | 525±25 | $0.2511 \pm 0.0239$ |
| 70 | ~5000 | 108 | $4750 \pm 250$ | 108 | $0.0456 \pm 0.0024$ |
| 71 | >5000 | 218 | $5250 \pm 250$ | 218 | $0.0832 \pm 0.0040$ |
| 72 | >5000 | 231 | $5250 \pm 250$ | 231 | $0.0882 \pm 0.0042$ |
| 73 | >5000 | 237 | $5250 \pm 250$ | 237 | $0.0905 \pm 0.0043$ |
| 74 | >6000 | >270 | $6300 \pm 300$ | 284±14 | $0.0906 \pm 0.0088$ |
| 75 | ~10000 | 28 | $9500 \pm 500$ | 28 | $0.0059 \pm 0.0003$ |
| 76 | ~10000 | ≈100 | $9500 \pm 500$ | 100±5 | $0.0212 \pm 0.0022$ |
| 77 | ~10000 | ~2000 | $9500 \pm 500$ | $1900 \pm 100$ | $0.4022 \pm 0.0422$ |
| 78 | >10000 | ≈400 | $10500 \pm 500$ | $400 \pm 20$ | $0.0766 \pm 0.0075$ |
| 79 | >10000 | >1000 | $10500 \pm 500$ | $1050 \pm 50$ | $0.2009 \pm 0.0191$ |
| 80 | ≈16000 | >700 | $16000 \pm 800$ | $735 \pm 35$ | $0.0923 \pm 0.0090$ |
| 81 | >16000 | >600 | $16800 \pm 800$ | $630 \pm 30$ | $0.0753 \pm 0.0072$ |
| 82 | >50000 | >3000 | $52500 \pm 2500$ | $3150 \pm 150$ | $0.1206 \pm 0.0115$ |

Note: ≈ denotes "about", ~ denotes "nearly", and > denotes "above".

into three categories with specified possible ranges according to their descriptions: i) "nearly $x$" (possible range $0.95x \pm 0.05x$); ii) "about $x$" (possible range $1.00x \pm 0.05x$); iii) "over $x$" (possible range $1.05x \pm 0.05x$). The full list of the data records is shown in Table 1. All data of matchmaking fairs are collected from the websites shown in Table 2.

In our model, $n$ is an internal parameter needed to be measured. Because a news report (descried as an experiment below) generally includes only the total of participants and the number of successful matching pairs, the male–female or female–male ratio $\eta$ defined in (7) should be estimated first. Under the assumption $k_1 \leq k_2$, the lower bound of $\eta$ is $\eta_{\min} = 1$, and once the total of participants $K = k_1 + k_2$ and the number of matching couples $E$ is determined, the upper bound of $\eta$ in that experiment is known: $\eta \leq (K - E)/E$.

Let $N$ be the number of experiments, $\eta_{\max,i}$ be the upper bound of $\eta$ in the $i$th experiment, and $H = \{\eta_{\max,1}, \eta_{\max,2}, \ldots, \eta_{\max,N}\}$ be the set of all upper bounds. By processing $N = 82$ experiments in Table 1, we obtain $\min(H) = 2$ and $\max(H) = 365.67$. Consider the least square criterion for fitting the model and the experimental data

$$Q(n,\eta) = \sum_{i=1}^{N} (P_{e,i} - P_{t,i}(n,\eta))^2 \qquad (13)$$

where $P_{e,i}$ denotes the experimental data in the $i$th experiment and $P_{t,i}(n,\eta)$ denotes the corresponding theoretical value calculated by (8), and the reality that in a matchmaking fair the numbers of males and females would not differ over some extent. We narrow the range of $\eta$ to [1,2] and solve this optimization problem

$$\min Q(n,\eta), \text{ subject to } 1 \le \eta \le 2, 1 \le n \le 10^4, \eta \in \mathbb{R}, n \in \mathbb{Z}. \quad (14)$$

**Table 2.** Data sources of matchmaking fairs.

| Website no. | Website name of matchmaking fairs |
|---|---|
| 01 | http://www.cdb.org.cn/newsview.php?id = 6359 |
| 02 | http://sd.people.com.cn/n/2012/0827/c183718-17407663.html |
| 03 | http://news.carnoc.com/list/183/183765.html |
| 04 | http://bbs.tiexue.net/post2_5756757_1.html |
| 05 | http://www.shxb.net/html/20110516/20110516_278862.shtml |
| 06 | http://news.qq.com/a/20100511/000472.htm |
| 07 | http://bbs.ganxianw.com/thread-46316-1-1.html |
| 08 | http://www.wzrb.com.cn/article321273show.html |
| 09 | http://www.wccdaily.com.cn/epaper/hxdsb/html/2012-05/14/content_448239.htm |
| 10 | http://wbnews.sxrb.com/news/ty/1372966.html |
| 11 | http://www.nbmz.gov.cn/view.aspx?id = 16595&AspxAutoDetectCookieSupport = 1 |
| 12 | http://nb.people.com.cn/GB/200892/16491824.html |
| 13 | http://cq.cqwb.com.cn/NewsFiles/201203/25/921397.shtml |
| 14 | http://www.sc.chinanews.com.cn/my/data/html/201212/32619.html |
| 15 | http://www.16466.com/info_detail.htm?id = 36526 |
| 16 | http://www.ncnews.com.cn/ncxw/shxw/t20121112_943114.htm |
| 17 | http://news.wzsee.com/2012/0502/130061.html |
| 18 | http://news.hexun.com/2012-08-27/145162214.html |
| 19 | http://www.douban.com/group/topic/28145139/ |
| 20 | http://zhuanti.10yan.com/zt/other/sdcms/html/xqj2012/xiangqindongtai/1377.html |
| 21 | http://www.wlmqwb.com/3229/syzt/hdzt/seven/201007/t20100719_1287834.shtml |
| 22 | http://www.dpcm.cn/html/news/shehui/20121211/8a485b96f289e038.htm |
| 23 | http://www.zhaogejia.com/News/Show/166 |
| 24 | http://cq.cqnews.net/shxw/shwx/200909/t20090928_3635782.htm |
| 25 | http://a.jiaodong.net/jiaoyou/detail/?/20120717134715.htm |
| 26 | http://www.dllake.com/testurl/news/news.asp?id = 1874 |
| 27 | http://fj.qq.com/a/20120413/000073.htm?pgv_ref = aio2012&ptlang = 2052 |
| 28 | http://epaper.lnd.com.cn/html/bdcb/20110118/bdcb635730.html |
| 29 | http://news.zh853.com/NewsShow-22166.html |
| 30 | http://news.163.com/11/1123/08/7JHJ4PP100014AED.html |
| 31 | http://news.xinmin.cn/rollnews/2011/05/03/10539635.html |
| 32 | http://www.0523qq.com/forum.php?mod = viewthread&tid = 2779 |
| 33 | http://news.ycw.gov.cn/html/2012-04/28/content_15150376.htm |
| 34 | http://epaper.lnd.com.cn/html/bdcb/20110118/bdcb635730.html |
| 35 | http://www.cqwb.com.cn/NewsFiles/201005/30/20102930062910354716.shtml |
| 36 | http://bddsb.bandao.cn/data/20120827/html/53/content_2.html |
| 37 | http://www.zhaogejia.com/News/Show/150 |
| 38 | http://3g.3xgd.com/news/play.asp?NewsID = 80975 |
| 39 | http://wed.cnhan.com/hjb/2012-12-03/3900.html |
| 40 | http://xt.fangyuan365.com/article/List.asp?ID = 8708 |
| 41 | http://cq.cqnews.net/shxw/2012-11/12/content_21432170.htm |
| 42 | http://xt.fangyuan365.com/article/List.asp?ID = 11694 |
| 43 | http://roll.sohu.com/20120625/n346389451.shtml |
| 44 | http://www.ijxjj.com/article/article_12773.html |
| 45 | http://news.xinmin.cn/shehui/2013/02/16/18638248_2.html |
| 46 | http://www.sz120.com/xwdt/ynxw/22205/ |
| 47 | http://www.sc.xinhuanet.com/content/2012-02/06/content_24649397.htm |
| 48 | http://dqnews.zjol.com.cn/dqnews/system/2010/08/17/012525402.shtml |
| 49 | http://cheshang.16888.com/newsinfo/2011/1115/141264.html |

**Table 2.** Cont.

| Website no. | Website name of matchmaking fairs |
|---|---|
| 50 | http://bbs.heze.cc/thread-842865-1-1.html |
| 51 | http://www.8hy.org/hyjy/hy6240/1 |
| 52 | http://www.cdrb.com.cn/html/2012-04/03/content_1546492.htm |
| 53 | http://heilongjiang.dbw.cn/system/2013/02/16/054584150.shtml |
| 54 | http://www.e0734.com/2012/0502/90707.html |
| 55 | http://sz.tznews.cn/tzwb/html/2012-07/09/content_71285.htm |
| 56 | http://www.xtrb.cn/epaper/ncwb/html/2011-08/09/content_275667.htm |
| 57 | http://www.hukou365.com/cwbbs/forum/showtopic_tree.jsp?rootid = 194730 |
| 58 | http://news.163.com/10/0329/03/62TPSMMO000146BB.html |
| 59 | http://www.chinajilin.com.cn/content/2009-02/15/content_1495554.htm |
| 60 | http://sy.house.sina.com.cn/news/2011-12-27/114483993.shtml |
| 61 | http://news.dayoo.com/guangzhou/201205/02/73437_23554932.htm |
| 62 | http://cq.qq.com/a/20090824/000190.htm |
| 63 | http://news.qq.com/a/20111228/000342.htm |
| 64 | http://www.efu.com.cn/data/2011/2011-08-09/389729.shtml |
| 65 | http://ent.163.com/12/1203/13/8HQ885MN00032DGD.html |
| 66 | http://www.subaonet.com/html/society/2010426/3C95FFIB98JI5FC.html |
| 67 | http://wb.sznews.com/html/2011-11/07/content_1812730.htm |
| 68 | http://heilongjiang.dbw.cn/system/2012/04/23/053819817.shtml |
| 69 | http://news.cnnb.com.cn/system/2011/10/31/007128083.shtml |
| 70 | http://www.people.com.cn/GB/paper447/17168/1505082.html |
| 71 | http://news.timedg.com/2012-04/16/content_9577975.htm |
| 72 | http://www.gddgart.com/artcenter/html3asp/town3ship/dq2012714_2357.asp |
| 73 | http://epaper.oeeee.com/I/html/2012-11/12/content_1751538.htm |
| 74 | http://news.hsw.cn/system/2010/06/28/050547974.shtml |
| 75 | http://fj.sina.com.cn/news/s/2012-08-24/07186785.html |
| 76 | http://net.chinabyte.com/164/12210164.shtml |
| 77 | http://epaper.hljnews.cn/shb/html/2008-05/26/content_199685.htm |
| 78 | http://www.estour.gov.cn/news/lvyouxinwen/2011/815/1181583840H7H40DE3ADE501J93BG9.shtml |
| 79 | http://www.048100.com.cn/news/bdxw/2009-04-20/617.html |
| 80 | http://www.estour.gov.cn/news/lvyouxinwen/2011/815/1181583840H7H40DE3ADE501J93BG9.shtml |
| 81 | http://www.estour.gov.cn/news/lvyouxinwen/2011/815/1181583840H7H40DE3ADE501J93BG9.shtml |
| 82 | http://zjnews.zjol.com.cn/05zjnews/system/2009/03/30/015385573.shtml |

Finally we obtain the estimation $n = 15$.

Figure 3 shows the relationship between the experimental data and the analytical predictions of our model. The red curve and olive curve are obtained from (8). The parameters of red curve are $\eta = 1$, $n = 15$; the parameters of olive curve are $\max(H) = 365.67$, $n = 15$. According to (7), when $k_1$ is equal to the minimum 1, $\xi$ takes the maximum value 225. The error bars of ordinate $P$ of round dots represent the ranges of empirical data $P$ in Table 1. Because $k_1$ is unknown and $\xi$ is undetermined, the bound for $k_1$ in the $i$th experiment is $\min(k_{1,i}) = E_i \le k_{1,i} \le \max(k_{1,i}) = (k_{1,i} + k_{2,i})/2$, and the bound for corresponding $\xi_i$ is $15^2/\max(k_{1,i}) \le \xi_i \le 15^2/\min(k_{1,i})$. Therefore, the ranges of abscissa $\xi_i$ of round dots are relatively wide and the middle points lie in $\xi_i = (15^2/\max(k_{1,i}) + 15^2/\min(k_{1,i}))/2$.

Figure 3 also shows when $\xi$ is relatively small and corresponding $k_1$ is big, all empirical data are enclosed between two curves; when $\xi$ is relatively big and corresponding $k_1$ is small, some empirical data are enclosed between the two curves, but other empirical data lie above the red curve and the trend of the empirical data is opposite to the analytical predictions. The possible reasons are: on the one hand, organizers of some matchmaking fairs select only a few participants meeting their requirements from a large number of applicants, so a participant is easier to find the right man or woman; on the other hand, when the number of participants is small in a matchmaking fair, they understand the difficulty of finding an ideal object so compromise to a goodish choice. The two reasons above cause that the fewer the participants are, the higher the matching probability $P$ is. Based on these effects, the deviation of experimental data from the model is acceptable.

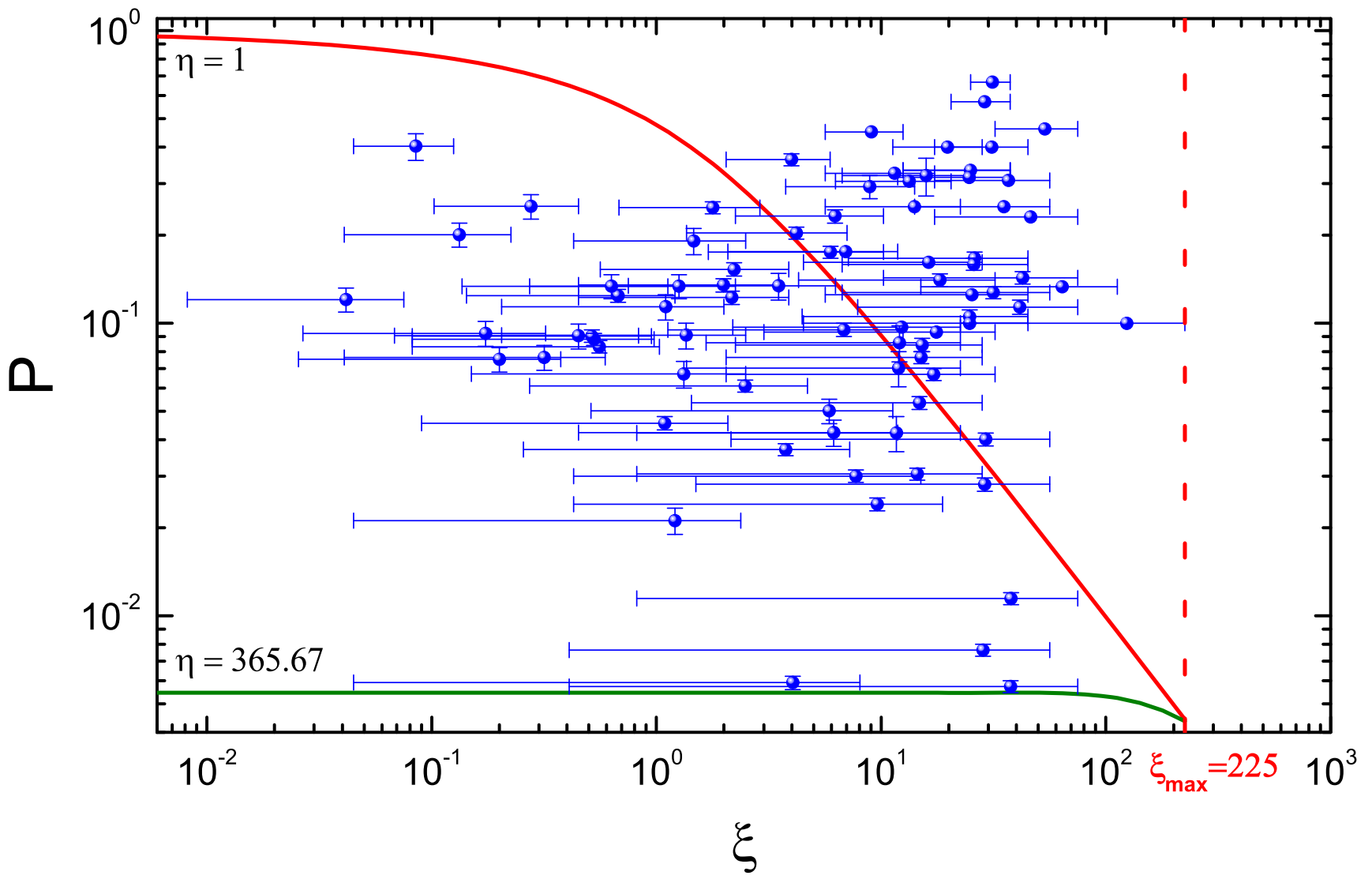

**Figure 3. The relationship between the experimental data and analytical predictions in the log-log plots.** The red curve and the olive curve are obtained from (8), and the parameters of red curve are $\eta = 1$, $n = 15$; The parameters of olive curve are $\eta = 365.67$, $n = 15$. The round dots represent the empirical data in Table 1. The $\xi_{max}$ represents the maximum value 225 of $\xi$.

## Conclusion

We propose a model of the two-way selection system and provide its analytical solution. Under several conditions, the compact approximations are derived analytically and verified by the simulation results. In the model, the parameter $n$ that denotes the number of characters directly determines the probability of the successful match – due to its importance, we propose a rough method to estimate its value by fitting the empirical data collected via the Internet and the result is $n = 15$. Under some artificial assumptions, most of the experimental data fall into the range predicted by our model, so this model is helpful for understanding the dynamics mechanism of the real-world two-way selection systems, and provides a starting point for researching the nature of real-world two-way selection systems. We believe our model could enlighten readers in this rapidly developing field.

## Author Contributions

Conceived and designed the experiments: BZ XH BW. Analyzed the data: BZ SQ XH JX. Wrote the paper: BZ SQ. Contributed analysis: ZH.